\documentclass{sig-alternate}

\usepackage{tikz}
\usetikzlibrary{arrows}
\usetikzlibrary{trees}
\usetikzlibrary{positioning}

\usepackage{array}
\usepackage{amstext}
\usepackage{mathtools}
\DeclarePairedDelimiter{\ceil}{\lceil}{\rceil}

\begin{document}

\title{IPFS - Content Addressed, Versioned, P2P File System (DRAFT 3)}
\subtitle{}

\numberofauthors{1}

\author{
\alignauthor
  Juan Benet\\
  \email{juan@benet.ai}
}

\maketitle
\begin{abstract}
The InterPlanetary File System (IPFS) is a peer-to-peer distributed file system that seeks to connect all computing devices with the same system of files. In some ways, IPFS is similar to the Web, but IPFS could be seen as a single BitTorrent swarm, exchanging objects within one Git repository. In other words, IPFS provides a high throughput content-addressed block storage model, with content-addressed hyper links. This forms a generalized Merkle DAG, a data structure upon which one can build versioned file systems, blockchains, and even a Permanent Web. IPFS combines a distributed hashtable, an incentivized block exchange, and a self-certifying namespace. IPFS has no single point of failure, and nodes do not need to trust each other.
\end{abstract}

\section{Introduction}

There have been many attempts at constructing a global distributed file system. Some systems have seen significant success, and others failed completely. Among the academic attempts, AFS~\cite{AFS} has succeeded widely and is still in use today. Others~\cite{Oceanstore, CFS} have not attained the same success. Outside of academia, the most successful systems have been peer-to-peer file-sharing applications primarily geared toward large media (audio and video). Most notably, Napster, KaZaA, and BitTorrent~\cite{BitTorrentUsers} deployed large file distribution systems supporting over 100 million simultaneous users. Even today, BitTorrent maintains a massive deployment where tens of millions of nodes churn daily~\cite{wang13}. These applications saw greater numbers of users and files distributed than their academic file system counterparts. However, the applications were not designed as infrastructure to be built upon. While there have been successful repurposings\footnote{For example, Linux distributions use BitTorrent to transmit disk images, and  Blizzard, Inc. uses it to distribute video game content.}, no general file-system has emerged that offers global, low-latency, and decentralized distribution.

Perhaps this is because a ``good enough'' system for most use cases already exists: HTTP.  By far, HTTP is the most successful ``distributed system of files'' ever deployed. Coupled with the browser, HTTP has had enormous technical and social impact. It has become the de facto way to transmit files across the internet. Yet, it fails to take advantage of dozens of brilliant file distribution techniques invented in the last fifteen years. From one prespective, evolving Web infrastructure is near-impossible, given the number of backwards compatibility constraints and the number of strong parties invested in the current model. But from another perspective, new protocols have emerged and gained wide use since the emergence of HTTP. What is lacking is upgrading design: enhancing the current HTTP web, and introducing new functionality without degrading user experience.

Industry has gotten away with using HTTP this long because moving small files around is relatively cheap, even for small organizations with lots of traffic. But we are entering a new era of data distribution with new challenges: (a) hosting and distributing petabyte datasets, (b) computing on large data across organizations, (c) high-volume high-definition on-demand or real-time media streams, (d) versioning and linking of massive datasets, (e) preventing accidental disappearance of important files, and more. Many of these can be boiled down to ``lots of data, accessible everywhere.'' Pressed by critical features and bandwidth concerns, we have already given up HTTP for different data distribution protocols. The next step is making them part of the Web itself.

Orthogonal to efficient data distribution, version control systems have managed to develop important data collaboration workflows. Git, the distributed source code version control system, developed many useful ways to model and implement distributed data operations. The Git toolchain offers versatile versioning functionality that large file distribution systems severely lack. New solutions inspired by Git are emerging, such as Camlistore~\cite{Camlistore}, a personal file storage system, and Dat~\cite{Dat} a data collaboration toolchain and dataset package manager. Git has already influenced distributed filesystem design~\cite{mashtizadeh13}, as its content addressed Merkle DAG data model enables powerful file distribution strategies. What remains to be explored is how this data structure can influence the design of high-throughput oriented file systems, and how it might upgrade the Web itself.

This paper introduces IPFS, a novel peer-to-peer version-controlled filesystem seeking to reconcile these issues. IPFS synthesizes learnings from many past successful systems. Careful interface-focused integration yields a system greater than the sum of its parts. The central IPFS principle is modeling \textit{all data} as part of the same Merkle DAG.

\section{Background}

This section reviews important properties of successful peer-to-peer systems, which IPFS combines.

\subsection{Distributed Hash Tables}

Distributed Hash Tables (DHTs) are widely used to coordinate and maintain metadata about peer-to-peer systems. For example, the BitTorrent MainlineDHT tracks sets of peers part of a torrent swarm.

\subsubsection{Kademlia DHT}

Kademlia~\cite{maymounkov02} is a popular DHT that provides:

\begin{enumerate}

  \item Efficient lookup through massive networks:
        queries on average contact $ \ceil{log_2 (n)} $ nodes.
        (e.g. $20$ hops for a network of $10,000,000$ nodes).

  \item Low coordination overhead: it optimizes the number of
        control messages it sends to other nodes.

  \item Resistance to various attacks by preferring long-lived nodes.

  \item Wide usage in peer-to-peer applications, including \\
        Gnutella and BitTorrent, forming networks of over 20 million nodes~\cite{wang13}.

 \end{enumerate}

\subsubsection{Coral DSHT}

While some peer-to-peer filesystems store data blocks directly in DHTs,
this ``wastes storage and bandwidth, as data must be stored at nodes where it
is not needed''~\cite{freedman04}. The Coral DSHT extends Kademlia in three
particularly important ways:

\begin{enumerate}

  \item Kademlia stores values in nodes whose ids are ``nearest'' (using
        XOR-distance) to the key. This does not take into account application
        data locality, ignores ``far'' nodes that may already have the data,
        and forces ``nearest'' nodes to store it, whether they need it or not.
        This wastes significant storage and bandwith. Instead, Coral stores
        addresses to peers who can provide the data blocks.

  \item Coral relaxes the DHT API from \texttt{get\_value(key)} to
        \texttt{get\_any\_values(key)} (the ``sloppy'' in DSHT).
        This still works since Coral users only need a single (working) peer,
        not the complete list. In return, Coral can distribute only subsets of
        the values to the ``nearest'' nodes, avoiding hot-spots (overloading
        \textit{all the nearest nodes} when a key becomes popular).

  \item Additionally, Coral organizes a hierarchy of separate DSHTs called
        \textit{clusters} depending on region and size. This enables nodes to
        query peers in their region first, ``finding nearby data without
        querying distant nodes''~\cite{freedman04} and greatly reducing the latency
        of lookups.

\end{enumerate}

\subsubsection{S/Kademlia DHT}

S/Kademlia~\cite{baumgart07} extends Kademlia to protect against malicious attacks in two particularly important ways:

\begin{enumerate}

  \item S/Kademlia provides schemes to secure \texttt{NodeId} generation,
        and prevent Sybill attacks. It requires nodes to create a PKI key pair, derive their identity from it, and sign their messages to each other. One scheme includes a proof-of-work crypto puzzle to make generating Sybills expensive.

  \item S/Kademlia nodes lookup values over disjoint paths, in order to
        ensure honest nodes can connect to each other in the presence of a large fraction of adversaries in the network. S/Kademlia achieves a success rate of 0.85 even with an adversarial fraction as large as half of the nodes.

\end{enumerate}

\subsection{Block Exchanges - BitTorrent}

BitTorrent~\cite{cohen03} is a widely successful peer-to-peer filesharing system, which succeeds in coordinating networks of untrusting peers (swarms) to cooperate in distributing pieces of files to each other. Key features from BitTorrent and its ecosystem that inform IPFS design include:

\begin{enumerate}
  \item BitTorrent's data exchange protocol uses a quasi tit-for-tat strategy
        that rewards nodes who contribute to each other, and punishes nodes who only leech others' resources.

  \item BitTorrent peers track the availability of file pieces, prioritizing
        sending rarest pieces first. This takes load off seeds, making non-seed peers capable of trading with each other.

  \item BitTorrent's standard tit-for-tat is vulnerable to some exploitative
        bandwidth sharing strategies. PropShare~\cite{levin08} is a different peer bandwidth allocation strategy that better resists exploitative strategies, and improves the performance of swarms.

\end{enumerate}

\subsection{Version Control Systems - Git}

Version Control Systems provide facilities to model files changing over time and distribute different versions efficiently. The popular version control system Git provides a powerful Merkle DAG \footnote{Merkle Directed Acyclic Graph -- similar but more general construction than a Merkle Tree. Deduplicated, does not need to be balanced, and non-leaf nodes contain data.} object model that captures changes to a filesystem tree in a distributed-friendly way.

\begin{enumerate}
  \item Immutable objects represent Files (\texttt{blob}), Directories (\texttt{tree}), and Changes (\texttt{commit}).
  \item Objects are content-addressed, by the cryptographic hash of their contents.
  \item Links to other objects are embedded, forming a Merkle DAG. This
  provides many useful integrity and workflow properties.
  \item Most versioning metadata (branches, tags, etc.) are simply pointer references, and thus inexpensive to create and update.
  \item Version changes only update references or add objects.
  \item Distributing version changes to other users is simply transferring objects and updating remote references.
\end{enumerate}

\subsection{Self-Certified Filesystems - SFS}

SFS~\cite{mazieres98, mazieres00} proposed compelling implementations of both (a) distributed trust chains, and (b) egalitarian shared global namespaces. SFS introduced a technique for building \textit{Self-Certified Filesystems}: addressing remote filesystems using the following scheme

\begin{verbatim}
      /sfs/<Location>:<HostID>
\end{verbatim}

\noindent where \texttt{Location} is the server network address, and:

\begin{verbatim}
      HostID = hash(public_key || Location)
\end{verbatim}

Thus the \textit{name} of an SFS file system certifies its server. The user can verify the public key offered by the server, negotiate a shared secret, and secure all traffic. All SFS instances share a global namespace where name allocation is cryptographic, not gated by any centralized body.

\section{IPFS Design}

IPFS is a distributed file system which synthesizes successful ideas from previous peer-to-peer sytems, including DHTs, BitTorrent, Git, and SFS. The contribution of IPFS is simplifying, evolving, and connecting proven techniques into a single cohesive system, greater than the sum of its parts. IPFS presents a new platform for writing and deploying applications, and a new system for distributing and versioning large data. IPFS could even evolve the web itself.

IPFS is peer-to-peer; no nodes are privileged. IPFS nodes store IPFS objects in local storage. Nodes connect to each other and transfer objects. These objects represent files and other data structures. The IPFS Protocol is divided into a stack of sub-protocols responsible for different functionality:

\begin{enumerate}
  \item \textbf{Identities} - manage node identity generation and verification. Described in Section 3.1.

  \item \textbf{Network} - manages connections to other peers, uses various underlying network protocols. Configurable. Described in Section 3.2.

  \item \textbf{Routing} - maintains information to locate specific peers and objects. Responds to both local and remote queries. Defaults to a DHT, but is swappable. Described in Section 3.3.

  \item \textbf{Exchange} - a novel block exchange protocol (BitSwap) that governs efficient block distribution. Modelled as a market, weakly incentivizes data replication. Trade Strategies swappable. Described in Section 3.4.

  \item \textbf{Objects} - a Merkle DAG of content-addressed immutable objects with links. Used to represent arbitrary datastructures, e.g. file hierarchies and communication systems. Described in Section 3.5.

  \item \textbf{Files} - versioned file system hierarchy inspired by Git. Described in Section 3.6.

  \item \textbf{Naming} - A self-certifying mutable name system. Described in Section 3.7.
\end{enumerate}

These subsystems are not independent; they are integrated and leverage
blended properties. However, it is useful to describe them separately,
building the protocol stack from the bottom up.

Notation: data structures and functions below are specified in Go syntax.

\subsection{Identities}

Nodes are identified by a \texttt{NodeId}, the cryptographic hash\footnote{Throughout this document, \textit{hash} and \textit{checksum} refer specifically to cryptographic hash checksums of data.} of a public-key, created with S/Kademlia's static crypto puzzle~\cite{baumgart07}. Nodes store their public and private keys (encrypted with a passphrase). Users are free to instatiate a ``new'' node identity on every launch, though that loses accrued network benefits. Nodes are incentivized to remain the same.

\begin{verbatim}
      type NodeId Multihash
      type Multihash []byte
      // self-describing cryptographic hash digest

      type PublicKey []byte
      type PrivateKey []byte
      // self-describing keys

      type Node struct {
        NodeId NodeID
        PubKey PublicKey
        PriKey PrivateKey
      }
\end{verbatim}

S/Kademlia based IPFS identity generation:

\begin{verbatim}
      difficulty = <integer parameter>
      n = Node{}
      do {
        n.PubKey, n.PrivKey = PKI.genKeyPair()
        n.NodeId = hash(hash(n.PubKey))
        p = count_preceding_zero_bits(n.NodeId)
      } while (p < difficulty)
\end{verbatim}

Upon first connecting, peers exchange public keys, and check: \texttt{hash(other.PublicKey) equals other.NodeId}. If not, the connection is terminated.

\paragraph{Note on Cryptographic Functions} Rather than locking the system to a particular set of function choices, IPFS favors self-describing values. Hash digest values are stored in \texttt{multihash} format, which includes a short header specifying the hash function used, and the digest length in bytes. Example:

\begin{verbatim}
    <function code><digest length><digest bytes>
\end{verbatim}

This allows the system to (a) choose the best function for the use case (e.g. stronger security vs faster performance), and (b) evolve as function choices change. Self-describing values allow using different parameter choices compatibly.

\subsection{Network}

IPFS nodes communicate regualarly with hundreds of other nodes in the network, potentially across the wide internet. The IPFS network stack features:

\begin{itemize}
  \item \textbf{Transport:} IPFS can use any transport protocol, and is best suited for WebRTC DataChannels~\cite{WebRTC} (for browser connectivity) or uTP(LEDBAT~\cite{LEDBAT}).
  \item \textbf{Reliability:} IPFS can provide reliability if underlying networks do not provide it, using uTP (LEDBAT~\cite{LEDBAT}) or SCTP~\cite{SCTP}.
  \item \textbf{Connectivity:} IPFS also uses the ICE NAT traversal techniques \cite{ICE}.
  \item \textbf{Integrity:} optionally checks integrity of messages using a hash checksum.
  \item \textbf{Authenticity:} optionally checks authenticity of messages using HMAC with sender's public key.
\end{itemize}

\subsubsection{Note on Peer Addressing}

IPFS can use any network; it does not rely on or assume access to IP. This allows IPFS to be used in overlay networks. IPFS stores addresses as \texttt{multiaddr} formatted byte strings for the underlying network to use. \texttt{multiaddr} provides a way to express addresses and their protocols, including support for encapsulation. For example:

\begin{verbatim}
  # an SCTP/IPv4 connection
  /ip4/10.20.30.40/sctp/1234/

  # an SCTP/IPv4 connection proxied over TCP/IPv4
  /ip4/5.6.7.8/tcp/5678/ip4/1.2.3.4/sctp/1234/
\end{verbatim}

\subsection{Routing}

IPFS nodes require a routing system that can find (a) other peers' network addresses and (b) peers who can serve particular objects. IPFS achieves this using a DSHT based on S/Kademlia and Coral, using the properties discussed in 2.1. The size of objects and use patterns of IPFS are similar to Coral \cite{freedman04} and Mainline~\cite{wang13}, so the IPFS DHT makes a distinction for values stored based on their size. Small values (equal to or less than \texttt{1KB}) are stored directly on the DHT. For values larger, the DHT stores references, which are the \texttt{NodeIds} of peers who can serve the block.

The interface of this DSHT is the following:

\begin{verbatim}

  type IPFSRouting interface {

    FindPeer(node NodeId)
    // gets a particular peer's network address

    SetValue(key []bytes, value []bytes)
    // stores a small metadata value in DHT

    GetValue(key []bytes)
    // retrieves small metadata value from DHT

    ProvideValue(key Multihash)
    // announces this node can serve a large value

    FindValuePeers(key Multihash, min int)
    // gets a number of peers serving a large value
  }
\end{verbatim}

Note: different use cases will call for substantially different routing systems (e.g. DHT in wide network, static HT in local network). Thus the IPFS routing system can be swapped for one that fits users' needs. As long as the interface above is met, the rest of the system will continue to function.

\subsection{Block Exchange - BitSwap Protocol}

In IPFS, data distribution happens by exchanging blocks with peers using a
BitTorrent inspired protocol: BitSwap. Like BitTorrent, BitSwap peers are
looking to acquire a set of blocks (\texttt{want\_list}), and have another set
of blocks to offer in exchange (\texttt{have\_list}).
Unlike BitTorrent, BitSwap is not limited to the blocks in one torrent.
BitSwap operates as a persistent marketplace where node can acquire the
blocks they need, regardless of what files those blocks are part of. The
blocks could come from completely unrelated files in the filesystem.
Nodes come together to barter in the marketplace.

While the notion of a barter system implies a virtual currency could be
created, this would require a global ledger to track ownership
and transfer of the currency. This can be implemented as a BitSwap Strategy, and will be explored in a future paper.

In the base case, BitSwap nodes have to provide direct value to each other
in the form of blocks. This works fine when the distribution of blocks across
nodes is complementary, meaning they have what the other wants. Often, this
will not be the case. In some cases, nodes must \textit{work} for their
blocks. In the case that a node has nothing that its peers want (or
nothing at all), it seeks the pieces its peers want, with lower
priority than what the node wants itself. This incentivizes nodes to cache and
disseminate rare pieces, even if they are not interested in them directly.

\subsubsection{BitSwap Credit}

The protocol must also incentivize nodes to seed when they do not need
anything in particular, as they might have the blocks others want. Thus,
BitSwap nodes send blocks to their peers optimistically, expecting the debt to
be repaid. But leeches (free-loading nodes that never share) must be protected against. A simple credit-like system solves the problem:

\begin{enumerate}
  \item Peers track their balance (in bytes verified) with other nodes.
  \item Peers send blocks to debtor peers probabilistically, according to
        a function that falls as debt increases.
\end{enumerate}

Note that if a node decides not to send to a peer, the node subsequently
ignores the peer for an \texttt{ignore\_cooldown} timeout. This prevents
senders from trying to game the probability by just causing more dice-rolls.
(Default BitSwap is 10 seconds).

\subsubsection{BitSwap Strategy}

The differing strategies that BitSwap peers might employ have wildly different effects on the performance of the exchange as a whole. In BitTorrent, while a standard strategy is specified (tit-for-tat), a variety of others have been implemented, ranging from BitTyrant~\cite{levin08} (sharing the least-possible), to BitThief~\cite{levin08} (exploiting a vulnerability and never share), to PropShare~\cite{levin08} (sharing proportionally). A range of strategies (good and malicious) could similarly be implemented by BitSwap peers. The choice of function, then, should aim to:

\begin{enumerate}
  \item maximize the trade performance for the node, and the whole exchange
  \item prevent freeloaders from exploiting and degrading the exchange
  \item be effective with and resistant to other, unknown
  strategies
  \item be lenient to trusted peers
\end{enumerate}

The exploration of the space of such strategies is future work.
One choice of function that works in practice is a sigmoid, scaled by a
\textit{debt retio}:

Let the \textit{debt ratio} $ r $ between a node and its peer be:
  \[ r = \dfrac{\texttt{bytes\_sent}}{\texttt{bytes\_recv} + 1} \]

Given $r$, let the probability of sending to a debtor be:
  \[ P\Big( \; send \; | \; r \;\Big) = 1 - \dfrac{1}{1 + exp(6-3r)} \]

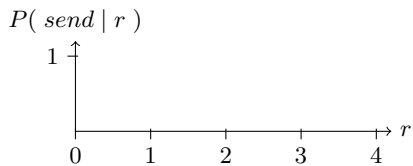
\begin{figure}
\centering
\begin{tikzpicture}[domain=0:4]

    \draw[->] (-0,0) -- (4.2,0) node[right] {$r$};
    \draw[->] (0,-0) -- (0,1.20) node[above] {$P(\;send\;|\;r\;)$};

    %ticks
    \foreach \x in {0,...,4}
      \draw (\x,1pt) -- (\x,-3pt)
        node[anchor=north] {\x};

    \foreach \y in {1,...,1}
      \draw (1pt,\y) -- (-3pt,\y)
        node[anchor=east] {\y};

    \draw[color=red] plot[] function{1 - 1/(1+exp(6-3*x))};

\end{tikzpicture}
\caption{Probability of Sending as $r$ increases}
\label{fig:psending-graph}
\end{figure}

As you can see in Figure \ref{fig:psending-graph}, this function drops off quickly as the nodes'
\textit{debt ratio} surpasses twice the established credit.
The \textit{debt ratio} is a measure of trust:
lenient to debts between nodes that have previously exchanged lots of data
successfully, and merciless to unknown, untrusted nodes. This
(a) provides resistance to attackers who would create lots of new nodes
(sybill attacks),
(b) protects previously successful trade relationships, even if one of the
nodes is temporarily unable to provide value, and
(c) eventually chokes relationships that have deteriorated until they
improve.

% \begin{center}
% \begin{tabular}{ >{$}c<{$} >{$}c<{$}}
%   P(\;send\;|\quad r) \;\;\;\;\;&  \\
%   \hline
%   \hline
%   P(\;send\;|\;0.0) =& 1.00 \\
%   P(\;send\;|\;0.5) =& 1.00 \\
%   P(\;send\;|\;1.0) =& 0.98 \\
%   P(\;send\;|\;1.5) =& 0.92 \\
%   P(\;send\;|\;2.0) =& 0.73 \\
%   P(\;send\;|\;2.5) =& 0.38 \\
%   P(\;send\;|\;3.0) =& 0.12 \\
%   P(\;send\;|\;3.5) =& 0.03 \\
%   P(\;send\;|\;4.0) =& 0.01 \\
%   P(\;send\;|\;4.5) =& 0.00 \\

% \end{tabular}
% \end{center}

% TODO look into computing share of the bandwidth, as described in propshare.

\subsubsection{BitSwap Ledger}

BitSwap nodes keep ledgers accounting the transfers with other nodes. This allows nodes to keep track of history and avoid tampering. When activating a connection, BitSwap nodes exchange their ledger information. If it does not match exactly, the ledger is reinitialized from scratch, losing the accrued credit or debt.  It is possible for malicious nodes to purposefully ``lose'' the Ledger, hoping to erase debts. It is unlikely that nodes will have accrued enough debt to warrant also losing the accrued trust; however the partner node is free to count it as misconduct, and refuse to trade.

\begin{verbatim}
      type Ledger struct {
        owner      NodeId
        partner    NodeId
        bytes_sent int
        bytes_recv int
        timestamp  Timestamp
      }
\end{verbatim}

Nodes are free to keep the ledger history, though it is not necessary for
correct operation. Only the current ledger entries are useful. Nodes are
also free to garbage collect ledgers as necessary, starting with the less
useful ledgers: the old (peers may not exist anymore) and small.

\subsubsection{BitSwap Specification}

BitSwap nodes follow a simple protocol.

\begin{verbatim}
    // Additional state kept
    type BitSwap struct {
      ledgers map[NodeId]Ledger
      // Ledgers known to this node, inc inactive

      active map[NodeId]Peer
      // currently open connections to other nodes

      need_list []Multihash
      // checksums of blocks this node needs

      have_list []Multihash
      // checksums of blocks this node has
    }

    type Peer struct {
      nodeid NodeId
      ledger Ledger
      // Ledger between the node and this peer

      last_seen Timestamp
      // timestamp of last received message

      want_list []Multihash
      // checksums of all blocks wanted by peer
      // includes blocks wanted by peer's peers
    }

    // Protocol interface:
    interface Peer {
      open (nodeid :NodeId, ledger :Ledger);
      send_want_list (want_list :WantList);
      send_block (block :Block) -> (complete :Bool);
      close (final :Bool);
    }
\end{verbatim}

Sketch of the lifetime of a peer connection:
\begin{enumerate}
  \item Open: peers send \texttt{ledgers} until they agree.
  \item Sending: peers exchange \texttt{want\_lists} and \texttt{blocks}.
  \item Close: peers deactivate a connection.
  \item Ignored: (special) a peer is ignored (for the duration of a timeout)
        if a node's strategy avoids sending

\end{enumerate}

\paragraph{Peer.open(NodeId, Ledger)}

When connecting, a node initializes a connection with a
\texttt{Ledger}, either stored from a connection in the past or a new one
zeroed out. Then, sends an Open message with the \texttt{Ledger} to the peer.

Upon receiving an \texttt{Open} message, a peer chooses whether to activate
the connection. If -- acording to the receiver's \texttt{Ledger} -- the sender
is not a trusted agent (transmission below zero, or large outstanding debt) the
receiver may opt to ignore the request. This should be done probabilistically
with an \texttt{ignore\_cooldown} timeout, as to allow errors to be corrected
and attackers to be thwarted.

If activating the connection, the receiver initializes a Peer object with the
local version of the \texttt{Ledger} and sets the \texttt{last\_seen}
timestamp. Then, it compares the received
\texttt{Ledger} with its own. If they match exactly, the connections have
opened. If they do not match, the peer creates a new zeroed out
\texttt{Ledger} and sends it.

\paragraph{Peer.send\_want\_list(WantList)}

While the connection is open, nodes advertise their
\texttt{want\_list} to all connected peers. This is done (a) upon opening the
connection, (b) after a randomized periodic timeout, (c) after a change in
the \texttt{want\_list} and (d) after receiving a new block.

Upon receiving a \texttt{want\_list}, a node stores it. Then, it checks whether
it has any of the wanted blocks. If so, it sends them according to the
\textit{BitSwap Strategy} above.

\paragraph{Peer.send\_block(Block)}

Sending a block is straightforward. The node simply transmits the block of
data. Upon receiving all the data, the receiver computes the Multihash
checksum to verify it matches the expected one, and returns confirmation.

Upon finalizing the correct transmission of a block, the receiver moves the
block from \texttt{need\_list} to \texttt{have\_list}, and both the receiver
and sender update their ledgers to reflect the additional bytes transmitted.

If a transmission verification fails, the sender is either malfunctioning or
attacking the receiver. The receiver is free to refuse further trades. Note
that BitSwap expects to operate on a reliable transmission channel, so
transmission errors -- which could lead to incorrect penalization of an honest
sender -- are expected to be caught before the data is given to BitSwap.

\paragraph{Peer.close(Bool)}

The \texttt{final} parameter to \texttt{close} signals whether the intention
to tear down the connection is the sender's or not. If false, the receiver
may opt to re-open the connection immediatelty. This avoids premature
closes.

A peer connection should be closed under two conditions:
\begin{itemize}
  \item a \texttt{silence\_wait} timeout has expired without receiving any
        messages from the peer (default BitSwap uses 30 seconds).
        The node issues \texttt{Peer.close(false)}.
  \item the node is exiting and BitSwap is being shut down.
        In this case, the node issues \texttt{Peer.close(true)}.
\end{itemize}

After a \texttt{close} message, both receiver and sender tear down the
connection, clearing any state stored. The \texttt{Ledger} may be stored for
the future, if it is useful to do so.

\paragraph{Notes}

\begin{itemize}
  \item Non-\texttt{open} messages on an inactive connection should be ignored.
        In case of a \texttt{send\_block} message, the receiver may check
        the block to see if it is needed and correct, and if so, use it.
        Regardless, all such out-of-order messages trigger a
        \texttt{close(false)} message from the receiver to force
        re-initialization of the connection.
\end{itemize}

% TODO: Rate Limiting / Node Silencing

\subsection{Object Merkle DAG}

The DHT and BitSwap allow IPFS to form a massive peer-to-peer system for storing and distributing blocks quickly and robustly. On top of these, IPFS builds a Merkle DAG, a directed acyclic graph where links between objects are cryptographic hashes of the targets embedded in the sources. This is a generalization of the Git data structure. Merkle DAGs provide IPFS many useful properties, including:

\begin{enumerate}
  \item \textbf{Content Addressing:} all content is uniquely identified by its
        \texttt{multihash} checksum, \textbf{including links}.
  \item \textbf{Tamper resistance:} all content is verified with its checksum.
        If data is tampered with or corrupted, IPFS detects it.
  \item \textbf{Deduplication:} all objects that hold the exact same content
        are equal, and only stored once. This is particularly useful with
        index objects, such as git \texttt{trees} and \texttt{commits}, or common portions of data.
\end{enumerate}

The IPFS Object format is:

\begin{verbatim}

    type IPFSLink struct {
      Name string
      // name or alias of this link

      Hash Multihash
      // cryptographic hash of target

      Size int
      // total size of target
    }

    type IPFSObject struct {
      links []IPFSLink
      // array of links

      data []byte
      // opaque content data
    }

\end{verbatim}

The IPFS Merkle DAG is an extremely flexible way to store data. The only requirements are that object references be (a) content addressed, and (b) encoded in the format above. IPFS grants applications complete control over the data field; applications can use any custom data format they chose, which IPFS may not understand. The separate in-object link table allows IPFS to:

\begin{itemize}

  \item List all object references in an object. For example:
\begin{verbatim}
> ipfs ls /XLZ1625Jjn7SubMDgEyeaynFuR84ginqvzb
XLYkgq61DYaQ8NhkcqyU7rLcnSa7dSHQ16x 189458 less
XLHBNmRQ5sJJrdMPuu48pzeyTtRo39tNDR5 19441 script
XLF4hwVHsVuZ78FZK6fozf8Jj9WEURMbCX4 5286 template

<object multihash> <object size> <link name>
\end{verbatim}

  \item Resolve string path lookups, such as \texttt{foo/bar/baz}. Given an object, IPFS resolves the first path component to a hash in the object's link table, fetches that second object, and repeats with the next component. Thus, string paths can walk the Merkle DAG no matter what the data formats are.

  \item Resolve all objects referenced recursively:
\begin{verbatim}
> ipfs refs --recursive \
  /XLZ1625Jjn7SubMDgEyeaynFuR84ginqvzb
XLLxhdgJcXzLbtsLRL1twCHA2NrURp4H38s
XLYkgq61DYaQ8NhkcqyU7rLcnSa7dSHQ16x
XLHBNmRQ5sJJrdMPuu48pzeyTtRo39tNDR5
XLWVQDqxo9Km9zLyquoC9gAP8CL1gWnHZ7z
...
\end{verbatim}

\end{itemize}

A raw data field and a common link structure are the necessary components for constructing arbitrary data structures on top of IPFS. While it is easy to see how the Git object model fits on top of this DAG, consider these other potential data structures:
(a) key-value stores
(b) traditional relational databases
(c) Linked Data triple stores
(d) linked document publishing systems
(e) linked communications platforms
(f) cryptocurrency blockchains.
These can all be modeled on top of the IPFS Merkle DAG, which allows any of these systems to use IPFS as a transport protocol for more complex applications.

\subsubsection{Paths}

IPFS objects can be traversed with a string path API. Paths work as they do in traditional UNIX filesystems and the Web. The Merkle DAG links make traversing it easy. Note that full paths in IPFS are of the form:

\begin{verbatim}
  # format
  /ipfs/<hash-of-object>/<name-path-to-object>

  # example
  /ipfs/XLYkgq61DYaQ8NhkcqyU7rLcnSa7dSHQ16x/foo.txt
\end{verbatim}

The \texttt{/ipfs} prefix allows mounting into existing systems at a standard mount point without conflict (mount point names are of course configurable). The second path component (first within IPFS) is the hash of an object. This is always the case, as there is no global root. A root object would have the impossible task of handling consistency of millions of objects in a distributed (and possibly disconnected) environment. Instead, we simulate the root with content addressing. All objects are always accessible via their hash. Note this means that given three objects in path \texttt{<foo>/bar/baz}, the last object is accessible by all:

\begin{verbatim}
    /ipfs/<hash-of-foo>/bar/baz
    /ipfs/<hash-of-bar>/baz
    /ipfs/<hash-of-baz>
\end{verbatim}

\subsubsection{Local Objects}

IPFS clients require some \textit{local storage}, an external system
on which to store and retrieve local raw data for the objects IPFS manages.
The type of storage depends on the node's use case.
In most cases, this is simply a portion of disk space (either managed by
the native filesystem, by a key-value store such as leveldb~\cite{dean11}, or
directly by the IPFS client). In others, for example non-persistent caches,
this storage is just a portion of RAM.

Ultimately, all blocks available in IPFS are in some node's
\textit{local storage}. When users request objects, they are
found, downloaded, and stored locally, at least temporarily. This provides
fast lookup for some configurable amount of time thereafter.

\subsubsection{Object Pinning}

Nodes who wish to ensure the survival of particular objects can do so by
\texttt{pinning} the objects. This ensures the objects are kept in the node's
\textit{local storage}. Pinning can be done recursively, to pin down all
linked descendent objects as well. All objects pointed to are then stored
locally. This is particularly useful to persist files, including references.
This also makes IPFS a Web where links are \textit{permanent}, and Objects can
ensure the survival of others they point to.

\subsubsection{Publishing Objects}

IPFS is globally distributed. It is designed to allow the files of millions of users to coexist together. The DHT, with content-hash addressing, allows publishing objects in a fair, secure, and entirely distributed way. Anyone can publish an object by simply adding its key to the DHT, adding themselves as a peer, and giving other users the object's path. Note that Objects are essentially immutable, just like in Git. New versions hash differently, and thus are new objects. Tracking versions is the job of additional versioning objects.

\subsubsection{Object-level Cryptography}

IPFS is equipped to handle object-level cryptographic operations. An encrypted or signed object is wrapped in a special frame that allows encryption or verification of the raw bytes.

\begin{verbatim}
    type EncryptedObject struct {
      Object []bytes
      // raw object data encrypted

      Tag []bytes
      // optional tag for encryption groups
    }

    type SignedObject struct {
      Object []bytes
      // raw object data signed

      Signature []bytes
      // hmac signature

      PublicKey []multihash
      // multihash identifying key
    }
\end{verbatim}

Cryptographic operations change the object's hash, defining a different object. IPFS automatically verifies signatures, and can decrypt data with user-specified keychains. Links of encrypted objects are protected as well, making traversal impossible without a decryption key. It is possible to have a parent object encrypted under one key, and a child under another or not at all. This secures links to shared objects.

\subsection{Files}

IPFS also defines a set of objects for modeling a versioned filesystem on top of the Merkle DAG. This object model is similar to Git's:

\begin{enumerate}
  \item \texttt{block}: a variable-size block of data.
  \item \texttt{list}: a collection of blocks or other lists.
  \item \texttt{tree}: a collection of blocks, lists, or other trees.
  \item \texttt{commit}: a snapshot in the version history of a tree.
\end{enumerate}

I hoped to use the Git object formats exactly, but had to depart to introduce certain features useful in a distributed filesystem, namely (a) fast size lookups (aggregate byte sizes have been added to objects), (b) large file deduplication (adding a \texttt{list} object), and (c) embedding of \texttt{commits} into \texttt{trees}. However, IPFS File objects are close enough to Git that conversion between the two is possible. Also, a set of Git objects can be introduced to convert without losing any information (unix file permissions, etc).

Notation: File object formats below use JSON. Note that this structure is actually binary encoded using protobufs, though ipfs includes import/export to JSON.

\subsubsection{File Object: \texttt{blob}}

The \texttt{blob} object contains an addressable unit of data, and
represents a file. IPFS Blocks are like Git blobs or filesystem data blocks. They store the users' data. Note that IPFS files can be represented by both \texttt{lists} and \texttt{blobs}. Blobs have no links.

\begin{verbatim}
{
  "data": "some data here",
  // blobs have no links
}
\end{verbatim}

\subsubsection{File Object: \texttt{list}}

The \texttt{list} object represents a large or deduplicated file made up of
several IPFS \texttt{blobs} concatenated together. \texttt{lists} contain
an ordered sequence of \texttt{blob} or \texttt{list} objects.
In a sense, the IPFS \texttt{list} functions like a filesystem file with
indirect blocks. Since \texttt{lists} can contain other \texttt{lists}, topologies including linked lists and balanced trees are possible. Directed graphs where the same node appears in multiple places allow in-file deduplication. Of course, cycles are not possible, as enforced by hash addressing.

\begin{verbatim}
{
  "data": ["blob", "list", "blob"],
    // lists have an array of object types as data
  "links": [
    { "hash": "XLYkgq61DYaQ8NhkcqyU7rLcnSa7dSHQ16x",
      "size": 189458 },
    { "hash": "XLHBNmRQ5sJJrdMPuu48pzeyTtRo39tNDR5",
      "size": 19441 },
    { "hash": "XLWVQDqxo9Km9zLyquoC9gAP8CL1gWnHZ7z",
      "size": 5286 }
    // lists have no names in links
  ]
}
\end{verbatim}

\begin{figure}
\centering
\begin{tikzpicture}[->,>=stealth',auto,thick,
  minimum height=2em,minimum width=5em]

  \tikzstyle{ghost}=[rectangle,rounded corners=.8ex];
  \tikzstyle{block}=[rectangle,draw,fill=blue!20,rounded corners=.8ex];
  \tikzstyle{list}=[rectangle,draw,fill=cyan!20,rounded corners=.8ex];
  \tikzstyle{tree}=[rectangle,draw,fill=green!20,rounded corners=.8ex];
  \tikzstyle{commit}=[rectangle,draw,fill=magenta!20,rounded corners=.8ex];
  \tikzstyle{every path}=[draw]

  \node[commit] (ccc111) {ccc111};
  \node[tree]   (ttt111) [below=3em of ccc111] {ttt111};
  \node[tree]   (ttt222) [below left=3em and 3em of ttt111] {ttt222};
  \node[tree]   (ttt333) [below=3em of ttt111] {ttt333};
  \node[ghost]  (ghost1) [below right=3em and 3em of ttt111] {};
  \node[list]   (lll111) [below=3em of ttt333] {lll111};
  \node[block]  (bbb111) [below=3em of ttt222] {bbb111};
  \node[block]  (bbb222) [below right=3em and 3em of ttt333] {bbb222};
  \node[block]  (bbb333) [below left=3em and 3em of lll111] {bbb333};
  \node[block]  (bbb444) [below=3em of lll111] {bbb444};
  \node[block]  (bbb555) [below right=3em and 3em of lll111] {bbb555};

  \path[every node/.style={font=\sffamily\small}]
    (ccc111) edge[out=-90,in=90] (ttt111)
    (ttt111) edge[out=-90,in=90] (ttt222)
             edge[out=-90,in=90] (ttt333)
             to  [out=-90,in=90]  (ghost1)
             to  [out=-90,in=90] (bbb222)
    (ttt222) edge[out=-90,in=90] (bbb111)
    (ttt333) edge[out=-90,in=90] (lll111)
             edge[out=-90,in=90] (bbb222)
    (lll111) edge[out=-90,in=90] (bbb333)
             edge[out=-90,in=90] (bbb444)
             edge[out=-90,in=90] (bbb555)
  ;

\end{tikzpicture}
\caption{Sample Object Graph} \label{fig:sample-object-graph}

\begin{verbatim}
    > ipfs file-cat <ccc111-hash> --json
    {
      "data": {
        "type": "tree",
        "date": "2014-09-20 12:44:06Z",
        "message": "This is a commit message."
      },
      "links": [
        { "hash": "<ccc000-hash>",
          "name": "parent", "size": 25309 },
        { "hash": "<ttt111-hash>",
          "name": "object", "size": 5198 },
        { "hash": "<aaa111-hash>",
          "name": "author", "size": 109 }
      ]
    }

    > ipfs file-cat <ttt111-hash> --json
    {
      "data": ["tree", "tree", "blob"],
      "links": [
        { "hash": "<ttt222-hash>",
          "name": "ttt222-name", "size": 1234 },
        { "hash": "<ttt333-hash>",
          "name": "ttt333-name", "size": 3456 },
        { "hash": "<bbb222-hash>",
          "name": "bbb222-name", "size": 22 }
      ]
    }

    > ipfs file-cat <bbb222-hash> --json
    {
      "data": "blob222 data",
      "links": []
    }
\end{verbatim}
\caption{Sample Objects} \label{fig:sample-objects}
\end{figure}

\subsubsection{File Object: \texttt{tree}}

The \texttt{tree} object in IPFS is similar to Git's: it represents a
directory, a map of names to hashes. The hashes reference \texttt{blobs}, \texttt{lists}, other \texttt{trees}, or \texttt{commits}. Note that traditional path naming is already implemented by the Merkle DAG.

\begin{verbatim}
{
  "data": ["blob", "list", "blob"],
    // trees have an array of object types as data
  "links": [
    { "hash": "XLYkgq61DYaQ8NhkcqyU7rLcnSa7dSHQ16x",
      "name": "less", "size": 189458 },
    { "hash": "XLHBNmRQ5sJJrdMPuu48pzeyTtRo39tNDR5",
      "name": "script", "size": 19441 },
    { "hash": "XLWVQDqxo9Km9zLyquoC9gAP8CL1gWnHZ7z",
      "name": "template", "size": 5286 }
    // trees do have names
  ]
}
\end{verbatim}

\subsubsection{File Object: \texttt{commit}}

The \texttt{commit} object in IPFS represents a snapshot in the version history of any object. It is similar to Git's, but can reference any type of object. It also links to author objects.

\begin{verbatim}
{
  "data": {
    "type": "tree",
    "date": "2014-09-20 12:44:06Z",
    "message": "This is a commit message."
  },
  "links": [
    { "hash": "XLa1qMBKiSEEDhojb9FFZ4tEvLf7FEQdhdU",
      "name": "parent", "size": 25309 },
    { "hash": "XLGw74KAy9junbh28x7ccWov9inu1Vo7pnX",
      "name": "object", "size": 5198 },
    { "hash": "XLF2ipQ4jD3UdeX5xp1KBgeHRhemUtaA8Vm",
      "name": "author", "size": 109 }
  ]
}
\end{verbatim}

\subsubsection{Version control}

The \texttt{commit} object represents a particular snapshot in the version
history of an object. Comparing the objects (and children) of two
different commits reveals the differences between two versions of the
filesystem. As long as a single \texttt{commit} and all the children objects
it references are accessible, all preceding versions are retrievable and the
full history of the filesystem changes can be accessed. This falls out
of the Merkle DAG object model.

The full power of the Git version control tools is available to IPFS users. The object model is compatible, though not the same. It is possible to (a) build a version of the Git tools modified to use the IPFS object graph, (b) build a mounted FUSE filesystem that mounts an IPFS \texttt{tree} as a Git repo, translating Git filesystem read/writes to the IPFS formats.

\subsubsection{Filesystem Paths}

As we saw in the Merkle DAG section, IPFS objects can be traversed with a string path API. The IPFS File Objects are designed to make mounting IPFS onto a UNIX filesystem simpler. They restrict \texttt{trees} to have no data, in order to represent them as directories. And \texttt{commits} can either be represented as directories or hidden from the filesystem entirely.

\subsubsection{Splitting Files into Lists and Blob}

One of the main challenges with versioning and distributing large files is finding the right way to split them into independent blocks. Rather than assume it can make the right decision for every type of file, IPFS offers the following alternatives:

\begin{enumerate}
  \item[(a)] Use Rabin Fingerprints \cite{RabinFingerprints} as in LBFS \cite{LBFS} to pick suitable block boundaries.
  \item[(b)] Use the rsync \cite{rsync} rolling-checksum algorithm, to detect blocks that have changed between versions.
  \item[(c)] Allow users to specify block-splitting functions highly tuned for specific files.
\end{enumerate}

\subsubsection{Path Lookup Performance}

Path-based access traverses the object graph. Retrieving
each object requires looking up its key in the DHT,
connecting to peers, and retrieving its blocks. This is considerable
overhead, particularly when looking up paths with many components.
This is mitigated by:

\begin{itemize}
  \item \textbf{tree caching}: since all objects are hash-addressed, they
        can be cached indefinitely. Additionally, \texttt{trees} tend to be
        small in size so IPFS prioritizes caching them over \texttt{blobs}.
  \item \textbf{flattened trees}: for any given \texttt{tree}, a special
        \texttt{flattened tree} can be constructed to list all objects
        reachable from the \texttt{tree}. Names in the \texttt{flattened tree}
        would really be paths parting from the original tree, with slashes.
\end{itemize}

For example, \texttt{flattened tree} for \texttt{ttt111} above:

\begin{verbatim}
{
"data":
  ["tree", "blob", "tree", "list", "blob" "blob"],
"links": [
  { "hash": "<ttt222-hash>", "size": 1234
    "name": "ttt222-name" },
  { "hash": "<bbb111-hash>", "size": 123,
    "name": "ttt222-name/bbb111-name" },
  { "hash": "<ttt333-hash>", "size": 3456,
    "name": "ttt333-name" },
  { "hash": "<lll111-hash>", "size": 587,
    "name": "ttt333-name/lll111-name"},
  { "hash": "<bbb222-hash>", "size": 22,
    "name": "ttt333-name/lll111-name/bbb222-name" },
  { "hash": "<bbb222-hash>", "size": 22
    "name": "bbb222-name" }
] }
\end{verbatim}

\subsection{IPNS: Naming and Mutable State}

So far, the IPFS stack forms a peer-to-peer block exchange constructing a content-addressed DAG of objects. It serves to publish and retrieve immutable objects. It can even track the version history of these objects. However, there is a critical component missing: mutable naming. Without it, all communication of new content must happen off-band, sending IPFS links. What is required is some way to retrieve mutable state at \textit{the same path}.

It is worth stating why -- if mutable data is necessary in the end -- we worked hard to build up an \textit{immutable} Merkle DAG. Consider the properties of IPFS that fall out of the Merkle DAG: objects can be (a) retrieved via their hash, (b) integrity checked, (c) linked to others, and (d) cached indefinitely. In a sense:

\begin{center}
  Objects are \textbf{permanent}
\end{center}

\noindent These are the critical properties of a high-performance distributed system, where data is expensive to move across network links. Object content addressing constructs a web with (a) significant bandwidth optimizations, (b) untrusted content serving, (c) permanent links, and (d) the ability to make full permanent backups of any object and its references.

The Merkle DAG, immutable content-addressed objects, and Naming, mutable pointers to the Merkle DAG, instantiate a dichotomy present in many successful distributed systems. These include the Git Version Control System, with its immutable objects and mutable references; and Plan9 \cite{Plan9}, the distributed successor to UNIX, with its mutable Fossil \cite{Fossil} and immutable Venti \cite{Venti} filesystems. LBFS \cite{LBFS} also uses mutable indices and immutable chunks.

\subsubsection{Self-Certified Names}

Using the naming scheme from SFS~\cite{mazieres98, mazieres00} gives us a way to construct self-certified names, in a cryptographically assigned global namespace, that are mutable. The IPFS scheme is as follows.

\begin{enumerate}
  \item  Recall that in IPFS:

\begin{verbatim}
NodeId = hash(node.PubKey)
\end{verbatim}

  \item We assign every user a mutable namespace at:

\begin{verbatim}
/ipns/<NodeId>
\end{verbatim}

  \item A user can publish an Object to this path \textbf{Signed} by her private key, say at:

\begin{verbatim}
/ipns/XLF2ipQ4jD3UdeX5xp1KBgeHRhemUtaA8Vm/
\end{verbatim}

  \item When other users retrieve the object, they can check the signature matches the public key and NodeId. This verifies the authenticity of the Object published by the user, achieving mutable state retrival.

\end{enumerate}

Note the following details:

\begin{itemize}
  \item The \texttt{ipns} (InterPlanetary Name Space) separate prefix is to establish an easily recognizable distinction between \textit{mutable} and \textit{immutable} paths, for both programs and human readers.

  \item Because this is \textit{not} a content-addressed object, publishing it relies on the only mutable state distribution system in IPFS, the Routing system. The process is (1) publish the object as a regular immutable IPFS object, (2) publish its hash on the Routing system as a metadata value:

\begin{verbatim}
routing.setValue(NodeId, <ns-object-hash>)
\end{verbatim}

  \item Any links in the Object published act as sub-names in the namespace:
\end{itemize}

\begin{verbatim}
/ipns/XLF2ipQ4jD3UdeX5xp1KBgeHRhemUtaA8Vm/
/ipns/XLF2ipQ4jD3UdeX5xp1KBgeHRhemUtaA8Vm/docs
/ipns/XLF2ipQ4jD3UdeX5xp1KBgeHRhemUtaA8Vm/docs/ipfs
\end{verbatim}

\begin{itemize}
  \item it is advised to publish a \texttt{commit} object, or some other object with a version history, so that clients may be able to find old names. This is left as a user option, as it is not always desired.

\end{itemize}

Note that when users publish this Object, it cannot be published in the same way

\subsubsection{Human Friendly Names}

While IPNS is indeed a way of assigning and reassigning names, it is not very user friendly, as it exposes long hash values as names, which are notoriously hard to remember. These work for URLs, but not for many kinds of offline transmission. Thus, IPFS increases the user-friendliness of IPNS with the following techniques.

\paragraph{Peer Links}

As encouraged by SFS, users can link other users' Objects directly into their own Objects (namespace, home, etc). This has the benefit of also creating a web of trust (and supports the old Certificate Authority model):

\begin{verbatim}
# Alice links to bob Bob
ipfs link /<alice-pk-hash>/friends/bob /<bob-pk-hash>

# Eve links to Alice
ipfs link /<eve-pk-hash/friends/alice /<alice-pk-hash>

# Eve also has access to Bob
/<eve-pk-hash/friends/alice/friends/bob

# access Verisign certified domains
/<verisign-pk-hash>/foo.com
\end{verbatim}

\paragraph{DNS TXT IPNS Records}

If \texttt{/ipns/<domain>} is a valid domain name, IPFS
looks up key \texttt{ipns} in its \texttt{DNS TXT} records. IPFS
interprets the value as either an object hash or another IPNS path:

\begin{verbatim}
    # this DNS TXT record
    ipfs.benet.ai. TXT "ipfs=XLF2ipQ4jD3U ..."

    # behaves as symlink
    ln -s /ipns/XLF2ipQ4jD3U /ipns/fs.benet.ai
\end{verbatim}

\paragraph{Proquint Pronounceable Identifiers}

There have always been schemes to encode binary into pronounceable words. IPNS supports Proquint \cite{Proquint}. Thus:

\begin{verbatim}
    # this proquint phrase
    /ipns/dahih-dolij-sozuk-vosah-luvar-fuluh

    # will resolve to corresponding
    /ipns/KhAwNprxYVxKqpDZ
\end{verbatim}

\paragraph{Name Shortening Services}

Services are bound to spring up that will provide name shortening as a service, offering up their namespaces to users. This is similar to what we see today with DNS and Web URLs:

\begin{verbatim}
    # User can get a link from
    /ipns/shorten.er/foobar

    # To her own namespace
    /ipns/XLF2ipQ4jD3UdeX5xp1KBgeHRhemUtaA8Vm
\end{verbatim}

\subsection{Using IPFS}

IPFS is designed to be used in a number of different ways. Here are just some of the usecases I will be pursuing:

\begin{enumerate}
  \item As a mounted global filesystem, under \texttt{/ipfs} and \texttt{/ipns}.
  \item As a mounted personal sync folder that automatically versions, publishes, and backs up any writes.
  \item As an encrypted file or data sharing system.
  \item As a versioned package manager for \textit{all} software.
  \item As the root filesystem of a Virtual Machine.
  \item As the boot filesystem of a VM (under a hypervisor).
  \item As a database: applications can write directly to the Merkle DAG data model and get all the versioning, caching, and distribution IPFS provides.
  \item As a linked (and encrypted) communications platform.
  \item As an integrity checked CDN for large files (without SSL).
  \item As an encrypted CDN.
  \item On webpages, as a web CDN.
  \item As a new Permanent Web where links do not die.
\end{enumerate}

The IPFS implementations target:

\begin{enumerate}
  \item[(a)] an IPFS library to import in your own applications.
  \item[(b)] commandline tools to manipulate objects directly.
  \item[(c)] mounted file systems, using FUSE \cite{FUSE} or as kernel modules.
\end{enumerate}

\section{The Future}

The ideas behind IPFS are the product of decades of successful distributed systems research in academia and open source. IPFS synthesizes many of the best ideas from the most successful systems to date. Aside from BitSwap, which is a novel protocol, the main contribution of IPFS is this coupling of systems and synthesis of designs.

IPFS is an ambitious vision of new decentralized Internet infrastructure, upon which many different kinds of applications can be built. At the bare minimum, it can be used as a global, mounted, versioned filesystem and namespace, or as the next generation file sharing system. At its best, it could push the web to new horizons, where publishing valuable information does not impose hosting it on the publisher but upon those interested, where users can trust the content they receive without trusting the peers they receive it from, and where old but important files do not go missing. IPFS looks forward to bringing us toward the Permanent Web.

\section{Acknowledgments}

IPFS is the synthesis of many great ideas and systems. It would be impossible to dare such ambitious goals without standing on the shoulders of such giants. Personal thanks to David Dalrymple, Joe Zimmerman, and Ali Yahya for long discussions on many of these ideas, in particular: exposing the general Merkle DAG (David, Joe), rolling hash blocking (David), and s/kademlia sybill protection (David, Ali). And special thanks to David Mazieres, for his ever brilliant ideas.

\section{References TODO}

\bibliographystyle{abbrv}
\bibliography{ipfs-cap2pfs}
%\balancecolumns
%\subsection{References}
\end{document}